\documentclass{dcds-b}
\usepackage{amsmath}
\usepackage{epsfig} 
\usepackage{graphics} 

\usepackage{hyperref}

\def\currentvolume{7}
\def\currentissue{2}
\def\currentyear{2007}
\def\currentmonth{March}
\def\ppages{441--456}

\newtheorem{theorem}{Theorem}

\newtheorem{lemma}{Lemma}

\theoremstyle{definition}
\newtheorem{definition}{Definition}

\title[Attractor bifurcation of the SH equation]
{Attractor bifurcation and final patterns\\ Of the n-dimensional
and generalized \\Swift-Hohenberg Equations}

\author[M. Yari]{}

\subjclass{Primary: 35G25,37G35; Secondary: 35B40}
\keywords{Attractor bifurcation, Swift-Hohenberg equation, Pattern formation}

\email{myari@indiana.edu}

\begin{document}
\maketitle

\centerline{\scshape Masoud Yari}
\medskip
{\footnotesize
\centerline{Department of Mathematics}
\centerline{Indiana University, Rawles Hall}
\centerline{Bloomington, IN 47405, USA}
} 

\medskip

\centerline{(Communicated by Jie Shen)}
\medskip

\begin{abstract}
In this paper I will investigate the bifurcation and asymptotic
behavior of solutions of the Swift-Hohenberg  equation and the
generalized Swift-Hohenberg equation with the Dirichlet boundary
condition on a one-dimensional domain $(0,L)$. I will also study the
bifurcation and stability of patterns in the $n$-dimensional
Swift-Hohenberg equation with the odd-periodic and periodic boundary
conditions. It is shown that each equation bifurcates from the
trivial solution to an attractor $\mathcal A_\lambda$ when the
control parameter $\lambda$ crosses $\lambda _{c} $, the principal
eigenvalue of $(I+\Delta)^2$. The local behavior of solutions and
their bifurcation to an invariant set near higher eigenvalues are
analyzed as well.
\end{abstract}

\section{Introduction}
Pattern formation is an interesting phenomenon which is often
observed in physics and chemistry. A physical system when driven
sufficiently far from equilibrium tends to form geometric patterns
\cite{ch}. This phenomena is determined by nonlinear aspects of the
system under study.

To study how those patterns form and evolve is the subject of
``non-equilibrium physics''. Non-equilibrium phenomena include
Taylor-Couette flow, parametric waves, reaction-diffusion systems,
propagation of electromagnetic waves in certain types of media, and
convection. Convection is a widely-studied example of the dynamics
that can occur in a system under the influence of a constant,
homogeneous temperature gradient. The mechanism of convection is
responsible for many phenomena of great interest, such as cloud
formation, ocean currents, plate tectonics, and crystal growth.

In the study of spatial patterns an important role is played by
model equations. A model equation, while simpler than full system of
equations, captures most features which control the pattern
formation phenomenon of the system. Recently attention has been
drawn to the study of fourth-order model equations involving
bistable dynamics. The fourth order pattern forming equation is of
central importance \cite{f, fk}:
$$ \frac{du}{dt}+ \frac{\partial^4 u}{\partial x^4}+q\frac{\partial^2 u}{\partial x^2}+ u^3 -u = 0,$$
where $q$ measures the pattern forming tendency. An extensive study
of the stationary equation, known as the Symmetric Bistable (SBS)
equation, can be found in \cite{pt}. As discussed in the latter,
stationary equations of a variety of fourth order model equations
such as the extended Fisher-Kolmogorov (EFK) equation, the
Swift-Hohenberg (SH) equation, the Suspension Bridge equation,
Bretherton's equation, the nonlinear Schrodinger equation can be
scaled to the Symmetric Bistable (SBS) equation (or canonical
equation). This can be put in a different way too. In fact the SBS
equation

$$\frac{du}{dt}+ A\frac{\partial^4 u}{\partial x^4} + B \frac{\partial^2 u}{\partial x^2} + Cu + u^3 =
0, \qquad A, B, C > 0;$$ can be written as
$$\frac{1}{C} \frac{du}{dt}+ ({\frac{\partial^2}{\partial x^2} + \frac{B}{2A}I })^2 u +
(\frac{C}{A}-(\frac{B}{2A})^2) u + \frac{1}{C} u^3 = 0.$$ and the
latter can be scaled to the dimensionless form of the
Swift-Hohenberg equation which shall be the main focus of this
paper.

The Swift-Hohenberg (SH) equation
$$ u_t = -(\frac{\partial^2}{\partial x^2 }+ I)^2u +
\lambda u - u^3,~~~ x\in(0,L), ~~\lambda \in \mathbb{R}.$$ was
proposed in 1977 in connection with Rayleigh-B\'enard's convection.
Although the SH equation cannot be derived systematically from the
Boussinesq equations, it captures much of the observed physical
behavior and has now become a general tool used to investigate not
only Rayleigh-B\'enard convection, but also other pattern-forming
systems \cite{ch}.

In this paper we will study qualitatively the bifurcation problem of
the 1-dimensional Swift-Hohenberg (SH) equation and generalized
Swift-Hohenberg (GSH) equation with the Dirichlet boundary
condition. Some numerical results, compatible with our results
concerning the SH equation with the Dirichlet boundary condition,
can be found in \cite{pr}.

The main technical tool is the new
bifurcation theory developed recently by Ma and Wang \cite{mw}. The
theory is based on a new notion of bifurcation called attractor
bifurcation. The main theorem associated with the attractor
bifurcation states that under certain conditions and provided the
critical state is asymptotically stable, when the first eigenvalue
of the linearized equation crosses the imaginary axis, the system
bifurcates from a trivial steady state solution to an attractor with
dimension between $m-1$ and $m$, where $m$ is the algebraic
multiplicity of the first eigenvalue. Another important ingredient
of the analysis is the reduction of the equation to its central
manifold. This involves lengthy calculations and careful examination
of the nonlinear interaction of higher order. The key idea is to
derive an higher order non-degenerate approximation which is
sufficient for the bifurcation analysis.

The main results obtained in this article can be summarized as
follows. First, we have shown that as the parameter $\lambda$
crosses the first critical value $\lambda_c$, the Swift-Hohenberg
equation bifurcates from the trivial solution to an attractor
$\mathcal A_{\lambda}$, with a dimension between $m-1$ and $m$
depending on boundary conditions, where $m$ is the multiplicity of
the first eigenvalue of the linearized problem. We will determine
the critical number $\lambda_c$ precisely. Second, as an attractor,
the bifurcated attractor $\mathcal A_{\lambda}$ has asymptotic
stability in the sense that it attracts all solutions with initial
data in the phase space outside of the stable manifold, with
codimension $m$, of the trivial solution. Third, the 1-d SH equation
with the Dirichlet boundary condition or odd periodic boundary
condition bifurcates to exactly two steady state solutions; and with
the periodic boundary condition, to an attractor homeomorphic to
$S^1$. Finally, in $n$-dimensional $(n\le 3)$ case with the odd
periodic boundary condition, the number of steady state solutions
contained in the bifurcated attractor $\mathcal A_{\lambda}$, for
$\lambda
> \lambda_c$, has been given precisely; and with the periodic
boundary condition, it is shown that the bifurcated attractor
$\mathcal A_{\lambda}$ contains a torus $\mathbb{T}^n$ which
consists of steady state solutions.

It is worth mentioning that although the problem has certain
symmetry, the method used in the article is crucial for proving the
bifurcated object is precisely the attractor and for proving the
stability property of the bifurcated object.

The paper is organized as follows. In section 2, we recall the main
results of the attractor bifurcation and center manifold reduction
method. In Section 3, we state and prove our main theorems
concerning the bifurcation of the Swift-Hohenberg equation and the
generalized Swift-Hohenberg equation with the Dirichlet boundary
condition. In section four, the main results concerning the
$n$-dimensional Swift-Hohenberg equation with the odd-periodic and
periodic boundary condition shall be stated.

\section{Abstract Bifurcation Theory and Reduction Method}

\subsection{Attractor Bifurcation Theorem}\label{sc2.1}

\label{sc2} Here we shall recall some results of dynamic bifurcation
of abstract nonlinear evolution equations developed in \cite{mw}.
Also we will refer the reader to \cite{mw-b} for a comprehensive
study of the dynamic bifurcation theory developed by Ma and Wang.

Let $H$ and $H_1$ be two Hilbert spaces, and $H_1 \hookrightarrow H$
be a dense and compact inclusion. Consider the following nonlinear
evolution equation
\begin{align}
\label{eq2.1}
& \frac{du}{dt} = L_\lambda u +G(u,\lambda), \\
\label{eq2.2} & u(0) = u_0,
\end{align}
where $u: [0, \infty) \rightarrow H$ is the unknown function,
$\lambda \in \mathbb R$ is the system parameter, and
$L_\lambda:H_1 \rightarrow H$ are parameterized linear completely
continuous fields continuously depending on $\lambda\in \mathbb R$,
which satisfy
\begin{equation}
\label{eq2.3} \left\{\begin{aligned}
& L_\lambda = -A + B_\lambda \,\,\, \text{ a sectorial operator}, \\
& A:H_1 \rightarrow H \,\,\,\, \text{a linear homeomorphism}, \\
& B_\lambda :H_1\rightarrow H \,\,\text{the parameterized linear
compact operators.}
\end{aligned}\right.
\end{equation}
We can see that $L_\lambda$ generates an analytic semi-group
$\{e^{-tL_\lambda}\}_{t\geq 0}$ and then we can define fractional
power operators $L^\alpha_\lambda$ for any $0\leq \alpha \leq 1$
with domain $H_\alpha = D(L^\alpha_\lambda)$ such that $H_{\alpha_1}
\subset H_{\alpha_2}$ if $\alpha_2 < \alpha_1$, and $H_0=H$.

We now assume that the nonlinear terms $G(\cdot, \lambda):H_\alpha
\to H$ for some $0\leq \alpha < 1$ are a family of parameterized
$C^r$ bounded operators ($r\geq 1$) continuously depending on the
parameter $\lambda\in \mathbb R$, such that
\begin{equation}
\label{eq2.4}
G(u,\lambda) = o(\|u\|_{H_\alpha}), \quad \forall\,\, \lambda\in \mathbb R.
\end{equation}

For the linear operator $A$ we assume that there exists a real
eigenvalue sequence $\{\rho_k\} \subset \mathbb R$ and an
eigenvector sequence $\{e_k\}\subset H_1$, i.e.,
\begin{equation}
\label{eq2.5} \left\{\begin{aligned}
& Ae_k = \rho_ke_k, \\
& 0<\rho_1\le \rho_2\le \cdots, \\
& \rho_k\rightarrow \infty \,\,(k\rightarrow \infty)
\end{aligned}\right.
\end{equation}
where $\{e_k\}$ is an orthogonal basis of $H$.

For the compact operator $B_\lambda:H_1 \rightarrow H$, we also
assume that there is a constant $0<\theta<1$ such that
\begin{equation}
\label{eq2.6} B_\lambda :H_\theta \longrightarrow H
\,\,\text{bounded, $\forall$ $\lambda\in \mathbb R$.}
\end{equation}

We know that the operator $L=-A+B_\lambda$ satisfying (\ref{eq2.5})
and (\ref{eq2.6}) is a sectorial operator. It generates an analytic
semigroup $\{S_\lambda (t)\}_{t \geq 0}$. Then the solution of
(\ref{eq2.1}) and (\ref{eq2.2}) can be expressed as
$$
u(t,u_0) = S_\lambda(t)u_0, \qquad t\geq 0.
$$

\begin{definition}
\label{df2.1} A set $\Sigma \subset H$ is called an invariant set of
(\ref{eq2.1}) if $S(t) \Sigma = \Sigma$ for any $t\geq 0$. An
invariant set $\Sigma \subset H$ of (\ref{eq2.1}) is said to be an
attractor if $\Sigma$ is compact, and there exists a neighborhood $U
\subset H$ of $\Sigma$ such that for any $\varphi\in U$ we have
\begin{equation}\label{eq2.7}
\lim_{t \rightarrow \infty} \text{\rm dist}_H(u(t,\varphi),\Sigma)=
0.
\end{equation}
The largest open set $U$ satisfying (\ref{eq2.7}) is called the
basin of attraction of $\Sigma$.
\end{definition}

\begin{definition}
\label{df2.2}
\begin{enumerate}

\item We say that the equation (\ref{eq2.1}) bifurcates from
$(u,\lambda) = (0,\lambda_0)$ an invariant set $\Omega_\lambda$, if
there exists a sequence of invariant sets $\{\Omega_{\lambda_n}\}$
of (\ref{eq2.1}), $0 \notin \Omega_{\lambda_n}$ such that

\begin{eqnarray*}
& {\displaystyle\lim_{n\to \infty}} \lambda_n = \lambda_0, \\
& {\displaystyle\lim_{n\to \infty} \max_{x\in \Omega_{\lambda_n}}}
|x| =0.
\end{eqnarray*}
\item If the invariant sets $\Omega_\lambda$ are attractors of
(\ref{eq2.1}), then the bifurcation is called attractor bifurcation.

\item If $\Omega_\lambda$ are attractors and are homotopy
equivalent to an $m$-dimensional sphere $S^m$, then the bifurcation
is called $S^m$-attractor bifurcation.

\end{enumerate}

\end{definition}
The following dynamic bifurcation theorem for (\ref{eq2.1}) was
proved in \cite{mw}.

\begin{theorem}[Attractor Bifurcation Theorem ]
\label{abt}
Assume that (\ref{eq2.3})--(\ref{eq2.6}) hold. Let
the eigenvalues (counting multiplicity) of $L_\lambda $ be given by
$\beta_1(\lambda)$, $\beta_2(\lambda)$, $\cdots$,
$\beta_k(\lambda)$, $\cdots$ $\in \mathbb C.$ Suppose that
\begin{equation}
\label{eq2.8} Re\beta_i(\lambda)
\begin{cases}
\begin{aligned}
&<0 &&\text{if} &&&\lambda<\lambda_0, \\
&=0 &&\text{if} &&&\lambda=\lambda_0, \\
&>0 &&\text{if} &&&\lambda>\lambda_0,
\end{aligned}
\end{cases}
\qquad (1\leq i\leq m),
\end{equation}
\begin{equation}
\label{eq2.9} Re\beta_j(\lambda_0) <0 \qquad (m+1 \leq j).
\end{equation}
Let the eigenspace of $L_\lambda$ at $\lambda_0$ be
\[ E_0 =\bigcup_{i=1}^{m}
\left\{ u \in H_1 \mid (L_{\lambda_0}-\beta_i(\lambda_0))^k u =0
, k=1,2, \cdots \right\}. \] and $u=0$ be a locally asymptotically stable
equilibrium point of (\ref{eq2.1}) at $\lambda=\lambda_0$. Then the
following assertions hold.

\begin{enumerate}
\item The equation (\ref{eq2.1}) bifurcates from $(u,\lambda) =
(0,\lambda_0)$ to an attractor $\mathcal A_\lambda$ for
$\lambda>\lambda_0$, with $m-1\leq \dim \mathcal A_\lambda \leq m$,
which is connected if $m>1$.

\item The attractor $\mathcal A_\lambda$ is a limit of a sequence of
$m$-dimensional annulus $M_k$ with $M_{k+1}\subset M_k$; in
particular if $\mathcal A_\lambda$ is a finite simplicial complex,
then $\mathcal A_\lambda$ has the homotopy type of $S^{m-1}$.

\item For any $u_\lambda \in \mathcal A_\lambda$, $u_\lambda$ can be
expressed as
\[
u_\lambda = v_\lambda + o(\|v_\lambda\|_{H_1}), \quad v_\lambda\in
E_0. \]

\item If the number of equilibrium points of (\ref{eq2.1}) in
$\mathcal A_\lambda$ is finite, then we have the following index
formula
\begin{displaymath}
\sum_{u_i \in \mathcal A_\lambda} ind[-(L_{\lambda} + G) , u_i] =
\begin{cases}
\begin{aligned}
&2 &&\text{if} &&&m = odd, \\
&0 &&\text{if} &&&m= even.
\end{aligned}
\end{cases}
\end{displaymath}
\item If $u=0$ is globally stable for (\ref{eq2.1}) at
$\lambda=\lambda_0$, then for any bounded open set $U\subset H$ with
$0\in U$, there is an $\varepsilon>0$ such that as
$\lambda_0<\lambda<\lambda_0 +\varepsilon$, the attractor
$\mathcal{A}_\lambda$ bifurcated from $(0,\lambda_0)$ attracts
$U\setminus \Gamma$ in $H$, where $\Gamma$ is the stable manifold of
$u=0$ with codimension $m$. In particular, if (\ref{eq2.1}) has a
global attractor for all $\lambda$ near $\lambda_0$, then
$\varepsilon$ can be chosen independently of U.

\end{enumerate}

\end{theorem}

The following theorem is an immediate result of the attractor
bifurcation theorem:

\begin{theorem} [Pitchfork bifurcation]\label{pbt} If the first eigenvalue is simple, i.e.
$m=1$, then the bifurcated attractor $\mathcal A_\lambda$ consists
of exactly two points $u_1$ and $u_2$. Moreover, for any bounded
open set $U\subset H$ with $0\in U$ there is an $\varepsilon>0$ such
that as $\lambda_0<\lambda<\lambda_0 +\varepsilon$, $U$ can be
decomposed into two open sets $U_1 ^\lambda$ and $U_2 ^\lambda$
satisfying such that

\begin{enumerate}
\item $\bar{U}= \bar{U_1 ^\lambda}+\bar{U_2 ^\lambda}, \bar{U_1 ^\lambda} \bigcap
\bar{U_2 ^\lambda} = \emptyset$
and $0\in \partial U_1 ^\lambda \bigcap \partial U_2 ^\lambda$,
\item $u_i ^\lambda \in U_i ^\lambda(i=1,2)$, and
\item ${\displaystyle\lim_{t\rightarrow \infty}}||u(t,\varphi) - u_i ^\lambda||_H = 0$,\\
for any $\varphi \in U^{\lambda} _i ( i=1,2)$, where $u(t,\varphi)$
is the solution of (\ref{eq2.1}).
\end{enumerate}

\end{theorem}

The following theorem will be useful later.

\begin{theorem}[\cite{mw-b}] \label{s1b}
Let $v$ be a two-dimensional $C^r$
($r\ge1$) vector field given by
$$v_\lambda = \lambda x - G_k(x,\lambda)+ o(|x|^k),$$ where $x \in
\mathbb{R}^2$ , $G_k$ is a k-multilinear field, and $k=2m+1$ ($m\ge1$).
If $G_k$ satisfies
$$ C_1 |x|^{k+1}\le <G_k(x,\lambda), x> \le C_2 |x|^{k+1},$$ for some
constant $C_2>C_1>0$, then $v_\lambda$ bifurcates from
$(x,\lambda)=(0,0)$ on $ \lambda > 0$ to an attractor
$\Omega_\lambda$, which is homeomorphic to $S^1$. Moreover, one and
only one of the following is true:
\begin{enumerate}
\item $\Omega_\lambda$ is a periodic orbit.
\item $\Omega_\lambda$ consists of only singular points.
\item $\Omega_\lambda$ contains at most $2(k+1) = 4(m+1)$
singular points, and has $4N+n (N+n \ge 1)$ singular points. 2N
of which are saddle points, $2N$ of which are stable node points
(possibly degenerate), and $n$ of which have index zero.
\end{enumerate}
\end{theorem}

\subsection{Reduction Method} \label{sc2.2}

A useful tool in the study of bifurcation problems is the reduction
of the equation to its local center manifold. The idea is to project
the equation to a finite dimensional space after a change of basis.
An extensive study of the method of reduction to center manifold can
be found in \cite{mw-b}.

Consider the following non-linear evolution equation:
\begin{equation}
\left\{%
\begin{aligned}
& \frac{du}{dt} = L_\lambda u +G(u,\lambda), \\
& u(0) = u_0.
\end{aligned}%
\right.
\end{equation}
with $L_\lambda = -A + B_\lambda : H_1 \longrightarrow H$ being a
symmetric linear continuous field, $G(.,\lambda) :H_1
\longrightarrow H$ being a $C^\infty$, which can be expressed as
$$G(u,\lambda)=\sum_{n=k}^{\infty} G_n(u,\lambda),\qquad \text{for
some } k \ge 2$$ where $G_n :H_1\times...\times H_1 \longrightarrow
H$ is an n-multiple linear mapping, and
$G_n(u,\lambda)=G_n(u,...,u,\lambda)$.

Let $\beta_i(\lambda)$ and $e_i(\lambda)$ be the eigenvalue and
eigenvector of $L_\lambda$ respectively. Since $L_\lambda$ is
symmetric, $\beta_i(\lambda)$'s are real. Assume that
$e_i(\lambda)$'s form an orthogonal basis for the space $H$. Also,
assume the following conditions hold true:
\begin{equation}
\beta_i(\lambda) \begin{cases}
\begin{aligned}
&<0 &&\text{if} &&&\lambda<\lambda_0, \\
&=0 &&\text{if} &&&\lambda=\lambda_0, \\
&>0 &&\text{if} &&&\lambda>\lambda_0,
\end{aligned}
\end{cases}
\qquad (1\leq i\leq m),
\end{equation}
\begin{equation}
\begin{cases}
\begin{aligned}
&\beta_j(\lambda_0) >0 &&\text{for} &&&m<j \leq m+n, \\
&\beta_j(\lambda_0) <0 &&\text{for} &&&m+n < j.
\end{aligned}
\end{cases}
\end{equation}
then with a change of basis, the equation can be written in the new
basis (one can think of this as Fourier series). After projection to
the subspace generated by the first eigenvalue, the equation can be
reduced to the central manifold as follows:

\begin{equation} \label{ReducedEq}
\frac{dx}{dt}=J_{m \lambda}x + g(x,\lambda),
\end{equation}
where $ x= (x_1,...,x_m)^t $ , $J_{m \lambda}$ is the Jordan matrix
corresponding to the first m eigenvalues of $L_\lambda$, and $
g(x,\lambda)= (g_1(x,\lambda),...,g_m(x,\lambda))^t $ with $
g_i(x,\lambda)= <G(x+\phi(x,\lambda)),e_i> $. Here $\phi(x,\lambda)$
is the center manifold function near $\lambda_0$. Finally the above
equation can be rewritten as follows:

\begin{equation}
\frac{dx}{dt}=J_{m \lambda}x + \sum_{p=k}^{k+N-1} F_p(x) +
o(x^{N+k-1}),\qquad N \ge 1.
\end{equation}
This last equation is called the $N^{th}$-order approximation of the
(\ref{ReducedEq})

\section{Bifurcation of The SH and The GSH Equations}
\label{sc3}

\subsection{The SH and The GSH Equations} The following nonlinear equation

\begin{equation}\label{gsh}
\left\{%
\begin{aligned}
&u_t = -(\frac{\partial^2}{\partial x^2 }+ I)^2 u
+ \lambda u - u^3 + \mu u^2 ~,~~~ \lambda \in \mathbb{R}, x \in (0,L);\\
&u(0,t) = u(L,t)= u^{''}(0,t) = u^{''}(L,t) = 0;\\
&u(x,0)=u_0(x).
\end{aligned}%
\right.
\end{equation}
is known as the Swift-Hohenberg (SH) equation when $\mu = 0$, and as
the generalized Swift-Hohenberg (GSH) when $\mu > 0.$ The SH
equation, proposed in 1977 by Swift and Hohenberg \cite{sh}, has
been shown to be a useful tool in the study of a variety of
problems, such as Taylor-Couette flow \cite{hs,pm}, and in the study
of lasers \cite{lmn}. The GSH equation was proposed later in
connection with the study of localized patterns. Extensive numerical
and analytical studies have been done on both equations; for example
see \cite{ce,ch,bpa,pt} for the SH equation ; and for the GSH
equation see \cite{AGLR,BGL,HMBD,GL}.

The existence of solutions of (\ref{gsh}) is established in
\cite{pt} using a topological shooting method. Some bifurcation
analyses and some numerical simulations, when $0 < \lambda < 1$, are
conducted in \cite{pr}.

\subsection{Functional settings} Now we employ tools introduced in the
second section to discuss the bifurcation of the SH and GSH equation
with the Dirichlet condition. In an appropriate functional setting
the equation \ref{gsh} can be expressed in the following form
\begin{equation}\label{sh}
\left\{%
\begin{aligned}
& \frac{du}{dt} = L_\lambda u +G(u,\lambda), \\
& u(0) = u_0,
\end{aligned}%
\right.
\end{equation}
where the operators $L_\lambda$ and $G$ are defined as follows:
\begin{equation}\label{1d-op}
\begin{aligned}
&L_\lambda = -A + B_{\lambda} : H_1 \hookrightarrow H,\\
&A = (I+\frac{\partial^2 u}{\partial x^2})^2: H_1 \hookrightarrow H, \\
&B_{\lambda} = \lambda I : H_1 \hookrightarrow H,\\
&Gu = \mu u^2 - u^3 : H_1 \hookrightarrow H.\\
\end{aligned}
\end{equation}
with Hilbert spaces $H$ and $H_1$:
\begin{equation}\label{1d-space}
\begin{aligned}
&H_1 = \{u \in H^4(0,L) |~ u , u'' = 0 ~\text{ at }~ x = 0 , L \},\\
&H = L^2(0,L). \\
\end{aligned}
\end{equation}

Since $H_1$ is compactly imbedded in $H$, $ H_1 \hookrightarrow H$,
it is clear that $B: H_1 \hookrightarrow H$ is a compact operator.
With an easy calculation one can see the eigenvalues of $ A : H_1
\hookrightarrow H $ are $\lambda_n = (1-(\frac{n\pi}{L})^2)^2$;
hence, assuming $L\neq n \pi$ for any $n\in\mathbb{N}$, $ A $ is a
homeomorphism. Therefore, $\L_\lambda = -A+B $ is a completely
continuous field. When $L=n \pi$ for some integer $n$, $A+I$ will be
a homeomorphism. In this case, $\L_{\lambda}$ is still a completely
continuous field, for $\L_\lambda$ can be written as the sum of
$A+I$ and $(\lambda-1)I$.

\subsection{Bifurcation of The SH and The GSH equations}

In general the eigenvalues of $A=(I+\frac{\partial^2 }{\partial
x^2})^2: H_1 \hookrightarrow H$ are $\lambda_n =P(\frac{n\pi}{L})$,
where $P(x)= (1-x^2)^2$ (see Figure1), and its principal eigenvalue
is $\lambda_c=\min\{~P(\frac{n\pi}{L})~|~n \ge 1~\}$. Therefore,
depending on the value of $L$, we might get a different critical
value and a different final profile associated with it.

\begin{center}
\begin{figure}[h]\label{fig1}
\includegraphics[height=1.7in]{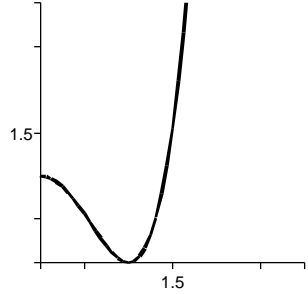}
\caption{\footnotesize The graph of $P(x)= (1-x^2)^2$ .}
\end{figure}
\end{center}

{\theorem \label{absh} Let $\lambda_c$ be the principal eigenvalue
of $(I+\frac{\partial^2 u}{\partial x^2})^2: H_1 \hookrightarrow H$.
Then the following assertions hold true for the SH equation with the
Dirichlet boundary condition (eq. \ref{gsh} with $\mu = 0$):
\begin{enumerate}
\item For $\lambda \le \lambda_c$, the trivial solution $u=0$ is globally asymptotically stable (Figure 2).
\item For $\lambda > \lambda_c$ The SH-d bifurcates from
$(0,\lambda_c)$ to an attractor bifurcation $\mathcal{A_{\lambda}}$
which consists of exactly two steady states (Figure 2).
\item For $\lambda>\lambda_c$, the bifurcated
attractor $\mathcal{A_{\lambda}}$ consists of exactly two steady
states ${u_1^{\lambda}, u_2^{\lambda}}$ given by
\begin{equation}\label{fpsh}
\left\{%
\begin{aligned}
&u_1^{\lambda} = \beta(\lambda) \phi_c + o(|\beta(\lambda)|), \\
&u_2^{\lambda} = - \beta(\lambda) \phi_c + o(|\beta(\lambda)|), \\
&\beta(\lambda) = \sqrt{\frac{\beta_c(\lambda)}{\alpha}}; \\
\end{aligned}%
\right.
\end{equation}
where $\beta_c(\lambda)$ is the first eigenvalue of the linearized
equation, $\phi_c$ the normalized eigenvector associated with it ,
and $\alpha = <(\phi_c)^3, \phi_c>_H$.

\item The stable manifold $\Gamma \subset H$ of $u=0$ separates
the phase space H into two open sets $U_{\lambda}^1 $and
$U_{\lambda}^2$, where $\lambda_c<\lambda<\lambda_c + \epsilon$ for
some $\epsilon>0$, which are the basin of attraction of
$u_1^{\lambda}$ and $u_2^{\lambda}$ respectively, i.e.
\begin{displaymath}
\begin{aligned}
&H = \overline{U^{\lambda}_1} + \overline{U^{\lambda}_2},\\
&U^{\lambda}_1 \cap U^{\lambda}_2 = \emptyset,\\
&\partial U^{\lambda}_1 \cap \partial U^{\lambda}_2 =
\Gamma,\\
&u_i^{\lambda} \in U^{\lambda}_i ~~~~ i =1,2,\\
&\lim_{t\rightarrow \infty}||u(t,\varphi) - u_i^{\lambda}||_H = 0,\\
\end{aligned}
\end{displaymath} for any $\varphi \in U^{\lambda}_i ( i=1,2)$,
where $u(t,\varphi)$ is the solution of the SH equation.

\item For any
integer $n$, the SH equation bifurcates
from $(u,\lambda)=(0,\lambda_n)$,on $\lambda > \lambda_n$,
to an attractor consisting of two steady state solutions of the SH equation.\\
\end{enumerate}}
For instance, when $L \le \pi$, the final patterns are given by
\begin{equation}\label{fpsh1}
\left\{%
\begin{aligned}
&u_1^{\lambda} = \beta(\lambda) \sin{\frac{\pi x}{L}} + o(|\beta(\lambda)|), \\
&u_2^{\lambda} = - \beta(\lambda) \sin{\frac{\pi x}{L}} + o(|\beta(\lambda)|), \\
&\beta(\lambda)= \sqrt{\frac{4}{3}(\lambda - (1 - (\frac{\pi}{L})^2)^2}. \\
\end{aligned}%
\right.
\end{equation}

\begin{center} {\setlength{\unitlength}{1mm}
\begin{picture}(60,65)
\thicklines
\put(-10,0){\vector(0,1){65}}
\put(-15, 30){\line(1,0){25}}
\put(10,30){\circle*{1}}
\put(13, 30){\line(1,0){3}}
\put(19, 30){\line(1,0){3}}
\put(25, 30){\line(1,0){3}}
\put(31, 30){\line(1,0){3}}
\put(37, 30){\line(1,0){3}}
\put(44,30){\vector(1,0){7}}
\put(1, 20){\line(0,1){20}}
\put(1,25){\vector(0,1){.5}}
\put(1,35){\vector(0,-1){.5}}
\put(1,30){\circle*{1}}
\qbezier(35.000,50.000)(-15.000,30.000)
(35.000,10.000)
\put(25, 0){\line(0,1){60}}
\put(25,20){\vector(0,-1){.5}}
\put(25,40){\vector(0,1){.5}}
\put(25,50){\vector(0,-1){.5}}
\put(25,10){\vector(0,1){.5}}
\put(25, 30){\circle{1}}
\put(25, 14.45){\circle*{1}}
\put(25, 45.3){\circle*{1}}
\put(26, 15){$u^{\lambda}_2$}
\put(26, 43.5){$u^{\lambda}_1$}
\put(-13,60){$u$}
\put(7,25){$\lambda_c$}
\put(48,25){$\lambda$}
\end{picture}}
\end{center}
\begin{center}
\vspace{.5cm} {\footnotesize Figure2. Pitchfork bifurcation of the
SH equation with the Dirichlet boundary condition.\\}
\end{center}
\vspace{.5cm}

\proof {\rm For the proof of the first part see Theorem (\ref{ga}). In fact the existence of the global
attractor for the SH equation is known \cite{pr}.

Without loss of generality, we will assume that $L<\pi$. The
eigenvectors and eigenvalues of $\L_\lambda : H_1 \hookrightarrow H$
are known to be:

\begin{equation}\label{evalue}
\begin{aligned}
&\beta_n(\lambda) = \lambda - (1-(\frac{n \pi}{L})^2)^2,\\
&\phi_n(x)=\sqrt{\frac{2}{L}}\sin(\frac{n \pi}{L} x).
\end{aligned}
\end{equation}

Moreover, the eigenvalues have the following properties:
\begin{displaymath}
\begin{aligned}
&\beta_{1}(\lambda) \left\{%
\begin{array}{ccc}
<0 & \text{if} & \lambda<\lambda_1, \\
=0& \text{if} & \lambda=\lambda_1,\\
>0 & \text{if} & \lambda>\lambda_1,\\
\end{array}%
\right.\\ &\beta_n(\lambda_1) < 0 \quad \forall n \neq 1.
\end{aligned}
\end{displaymath}

Hence this theorem is a direct result of the pitchfork bifurcation
theorem(\ref{pbt}). We only need to prove (\ref{fpsh1}). This can be
proven by the Lyapunov-Schmidt reduction method near $\lambda_c =
\lambda_1$. Let $ u \in H$ and $ u = \sum_{k=1}^{\infty}x_k
\phi_k(x)$. Then the steady state bifurcation equation of the SH
equation can be expressed as
$$\beta_n x_n - \sqrt{\frac{2}{L}}\int_{0}^{L} u^3 sin(\frac{n\pi
x}{L}) dx = 0,$$
where $\beta_n = \lambda - (1-(\frac{n \pi}{L})^2)^2 $. We have
\begin{align*}
u^3 = \sqrt{(\frac{2}{L})^3}\frac 14 \sum_{j,k,l \in \mathbb{N}} x_j x_k x_l
[\sin(j+k-l)\frac{\pi x}{L}+ \sin (j+l-k) \frac{\pi x}{L} +\\
\qquad \sin (k+l-j)\frac{\pi x}{L} - \sin(j+k+l)\frac{\pi x}{L} ];
\end{align*}
so we will have

\begin{displaymath}
\begin{aligned}
&x_1 = \frac{1}{2 L\beta_1} [ 3 \sum_{\substack{j \in \mathbb{N}\\k+l=1+j}} x_j x_k x_l ],\\
&x_2 = \frac{1}{2L\beta_2} [3 \sum_{\substack{j\in \mathbb{N}\\k+l=2+j}} x_j x_k x_l],\\
&x_3 = \frac{1}{2L\beta_3} [ 3 \sum_{\substack{j\in \mathbb{N}\\k+l=3+j}} x_j x_k x_l
- x_1 ^3],\\
&x_n = \frac{1}{2L\beta_n} [ 3 \sum_{\substack{j\in
\mathbb{N}\\k+l=n+j}} x_j x_k x_l -\sum_{\substack{j,k,l \in
\mathbb{N}\\j+k+l=n}} x_j x_k x_l],\quad n \ge 4.\\
\end{aligned}
\end{displaymath}
Hence, by induction, we have:
\begin{displaymath}
\begin{aligned}
&x_2 = o(|x_1| ^3 ),\\
&x_3 =-\frac{1}{2L\beta_3}x_1 ^3 + o(|x_1|^3 ),\\
&x_n= c_n x_1 ^n + o(|x_1| ^n),
\end{aligned}
\end{displaymath}
for $n \ge 4$ , where $c_n$ is a constant. Therefore, we have the
following bifurcation equation for the SH equation:
\begin{equation}\label{be}
2L \beta_1 x_1 - 3 x_1 ^3 + o(|x_1|^3 ) = 0.
\end{equation}
This completes the proof. The proof of the last part of the theorem
is similar.
\begin{flushright}
\vspace{-.8cm} $\Box$
\end{flushright}

For the generalized Swift-Hohenberg (GSH) equation, the stability of
the trivial solution $u=0$ may not be true at $\lambda_c$;
nevertheless, we can prove the existence of a bifurcation near the
critical value. {\theorem \label{abgsh} The following assertion hold
true for the GSH equation (eq. \ref{gsh} when $\mu
> 0$) with Dirichlet boundary conditions :
\begin{enumerate}
\item (\ref{gsh}) bifurcates from $(0,\lambda_c)$ to a unique
saddle point $u^{\lambda}$ (with Morse index one) on $\lambda <
\lambda_c$, and to a unique attractor $u^{\lambda}$ on $\lambda >
\lambda_c$.

\item If $\lambda > \lambda_c$ there is an open set $U$ of $u = 0$
which is divided into two open sets by the stable manifold $\Gamma$
of $u=0$ with codimension one in $H$.
\begin{displaymath}
\begin{aligned}
& \overline{U} = \overline{U^{\lambda}_1} + \overline{U^{\lambda}_2},\\
&U^{\lambda}_1 \cap U^{\lambda}_2 = \emptyset,\\
&\partial U^{\lambda}_1 \cap \partial U^{\lambda}_2 =
\Gamma,\\
&u^{\lambda} \in U^{\lambda}_1, and \\
&\lim_{t\rightarrow \infty}||u(t,\varphi) - u^{\lambda}||_H = 0,\\
\end{aligned}
\end{displaymath} for any $\varphi \in U^{\lambda}_1$,
where $u(t,\varphi)$ is the solution of (\ref{gsh}).

\item Near $\lambda_c$, the bifurcated
singular points $u_{\lambda}$ can be expressed as
$$ u^{\lambda}= \beta(\lambda) \phi_c + 0(|\beta(\lambda)|). $$
where $\beta_c(\lambda)$ is the first eigenvalue of the linearized
equation, $\phi_c$ the normalized eigenvector corresponding to it,
$\alpha = <(\phi_c)^2, \phi_c>_H$,
$\beta(\lambda) = -\frac{\beta_c}{\mu \alpha}$.\\
For instance when $L<\pi$, that is $\lambda_c=
(1-(\frac{\pi}{L})^2)^2$, we have $\alpha =
\frac{8\sqrt{2}}{3\pi\sqrt{L}}$ and
$$ u^{\lambda}= \frac{3\pi}{8\mu} (\lambda_1-\lambda) \sin(\frac{\pi}{L} x) + 0(|\lambda_1-\lambda|). $$
\end{enumerate}}
{\proof \rm By reducing the equation to its center manifold near
$\lambda_c$, we get the following equation when $\mu >0$ :

\begin{displaymath}
\frac{dx_c}{dt} = \beta_c x_1 + \alpha \mu x_c^2 + o(x_c^2).
\end{displaymath}
where $\alpha = <(\phi_c)^3, \phi_c>_H$ is a constant. Then the
theorem is an obvious result of this reduced equation.}
\begin{flushright}
\vspace{-.8cm} $\Box$
\end{flushright}

\setlength{\unitlength}{1mm}
\begin{picture}(40,45)
\thicklines

\put(0, 20){\line(1,0){40}}
\put(10,10){\line(0,1){20}}
\put(30,10){\line(0,1){20}}
\put (15,-4){$\lambda < \lambda_c$}
\put (11,17){$u_\lambda$}
\put (31,16){$0$}
\put(5,20){\vector(-1,0){.5}}
\put(20,20){\vector(1,0){.5}}
\put(35,20){\vector(-1,0){.5}}
\put(10,15){\vector(0,1){.5}}
\put(10,25){\vector(0,-1){.5}}
\put(30,15){\vector(0,1){.5}}
\put(30,25){\vector(0,-1){.5}}

\put(50, 20){\line(1,0){20}}
\put(60,10){\line(0,1){20}}
\put (55,-4){$\lambda = \lambda_c$}
\put (61,16){$0$}
\put(60,15){\vector(0,1){.5}}
\put(60,25){\vector(0,-1){.5}}
\put(55,20){\vector(-1,0){.5}}
\put(65,20){\vector(-1,0){.5}}

\put(80, 20){\line(1,0){40}}
\put(90,10){\line(0,1){20}}
\put(110,10){\line(0,1){20}}
\put(90,15){\vector(0,1){.5}}
\put(90,25){\vector(0,-1){.5}}
\put(110,15){\vector(0,1){.5}}
\put(110,25){\vector(0,-1){.5}}
\put(85,20){\vector(-1,0){.5}}
\put(100,20){\vector(1,0){.5}}
\put(115,20){\vector(-1,0){.5}}

\put (95,-4){$\lambda > \lambda_c$}
\put (91,16){$0$}
\put (111,17){$u_\lambda$}
\end{picture}

\vspace{1cm}
\begin{center}
{\footnotesize Figure3. Topological structure of dynamic bifurcation
of the GSH equation.
The horizonal line represents the center manifold.}
\end{center}

\section{ Bifurcation of Periodic Solutions of The Swift-Hohenberg equation}
\subsection{$n$-dimensional Swift-Hohenberg equation}

The $n$-dimensional Swift-Hohenberg equation reads as
\begin{equation}\label{SH-p}
\left \{%
\begin{aligned}
&\frac{du}{dt}= - (I+\Delta)^2 u + \lambda u - u^3, \quad x \in \Omega,\quad t > 0;\\
&u(0) = u_0;
\end{aligned}
\right.
\end{equation}
where $\Omega = (0 , L )^n$, $1 \le n \le 3$. And the equation is
supplemented with one of the following boundary conditions:
\begin{enumerate}
\item \textit{The odd-periodic boundary condition:}\\ $$u(x_i,t) = u(x_i+L_i,t)
\quad \text{and} \quad u(-x,t)= - u(x,t);$$
\item \textit{The periodic boundary condition:}\\ $$u(x_i,t) = u(x_i+L_i,t) \quad \forall i.$$
\end{enumerate}
\subsection{Functional settings}
In an appropriate functional setting the SH equation(\ref{SH-p}) can be expressed
in the following form
\begin{equation}\label{sh}
\left\{%
\begin{aligned}
& \frac{du}{dt} = L_\lambda u +G(u,\lambda), \\
& u(0) = u_0,
\end{aligned}%
\right.
\end{equation}
where the operators $L_\lambda$ and $G$ are defined as follows:
\begin{equation}\label{op}
\begin{aligned}
&L_\lambda = -A + B_{\lambda} : H_1 \hookrightarrow H,\\
&A = (I+\Delta)^2: H_1 \hookrightarrow H, \\
&B_{\lambda} = \lambda I : H_1 \hookrightarrow H,\\
&Gu = - u^3 : H_1 \hookrightarrow H.\\
\end{aligned}
\end{equation}
with Hilbert spaces $H$ and $H_1$:
\begin{equation}\label{spaces}
\begin{aligned}
& H_1 = \left\{%
\begin{aligned}
&\{u \in \dot{H}^4_{per}(\Omega) | u (-x,t) = -u(x,t)\} &&\text{for odd-periodic condition}, \\
& \dot{H}^4_{per}(\Omega) &&\text{for periodic condition}; \\
\end{aligned}%
\right.\\ & H = \left\{%
\begin{aligned}
&\{u \in \dot{L}^2 _{per}(\Omega) | u (-x,t) = -u(x,t)\} &&\text{for odd-periodic condition}, \\
& \dot{L}^2 _{per}(\Omega) &&\text{for periodic condition}. \\
\end{aligned}%
\right.\\
\end{aligned}
\end{equation}
The subscript "per" stands for "periodic" and the dot, $^.$ , means
$\int_0^L f dx = 0$, for $f$ in $H^4$ or $L^2$. In any case $H_1$ is
a dense and compact subspace of $H$, $H_1 \hookrightarrow H$.
{\theorem \label{ga}{\rm (An a priori estimate)} Assume $\lambda_c$ be the first eigenvalue
of $ (I+\Delta)^2: H_1 \hookrightarrow H$. Then the following a priori estimates hold for the SH equation:
\begin{equation}\label{abset1}
\hspace{0.6cm} |u(t)|_2 \le \begin{cases}
\begin{aligned}
& e^{(\lambda-\lambda_c)t} |u(0)| &&\text{if} &&&\lambda<\lambda_c; \\
& \frac{|u(0)|}{\sqrt{2 |\Omega| |u(0)|^2 t + 1}} &&\text{if} &&&\lambda=\lambda_c; \\
& \max \{ \sqrt{\frac{\lambda-\lambda_c}{|\Omega|}} , |u(0)| \}
&&\text{if} &&&\lambda>\lambda_c.
\end{aligned}
\end{cases}
\end{equation}
{\lemma \label{lemma1} Suppose $\psi: \mathbb{R}^+ \rightarrow
\mathbb{R}^+$ satisfies the following:
$$\psi'(t) \leq a \psi (t) - b \psi^2 (t),$$ with $b > 0$. Then the following assertions are true:\vspace{.2cm}
\begin{enumerate}
\item if $a<0$ then $\psi(t) \le \psi(0) e^{at}$,\vspace{.2cm}
\item if $a= 0$ then $\psi(t) \leq \frac{\psi(0)}{b \psi(0) t + 1},$ \vspace{.2cm}
\item if $a>0$ then $\psi(t) \le \max\{\psi(0), \sqrt{\frac ba}\}.$
\end{enumerate}}
\proof[proof of theorem \ref{ga}] {Using the energy method, from
(\ref{SH-p}) we will have:
$$ \frac12 \frac{d}{dt}|u|^2 _2 = - < (I+\Delta)^2(u) , u
> + \lambda_c |u|^2 _2 - |u^2|^2 _2.$$ It is known that the principal eigenvalue of
$A= (I+\Delta)^2: H_1 \hookrightarrow H $ satisfies $$\lambda_c = \min _ { \substack {u \in H_1 \\
u \neq 0}} \frac{<(I+\Delta)^2(u),u>}{|u|_{2}^{2}};$$ also by
H\"olders inequality :$$ |u|_{2}^{2} = \int_{\Omega} u^2 dx \leq
(|\Omega|)^\frac12 (\int_{\Omega} (u^2)^2)^\frac12 dx, $$ that is $$
- |u^2|^2 _2 \leq -\frac {1}{|\Omega|} |u|_{2}^{4}.$$ So we will
have
$$\frac{d}{dt}|u|^2 _2 \leq 2(\lambda_c - \lambda) |u|_2^2- \frac {2}{|\Omega|} |u|_{2}^{4}.$$
Therefore \ref{abset1} follows from the lemma(\ref{lemma1}).
\begin{flushright}
\vspace{-.8cm} $\Box$
\end{flushright}}

By the above theorem we know that $ u = 0 $ is globally
asymptotically stable for $ \lambda \le \lambda_c$. In fact, the
existence of the global attractor for the SH equation can be proven
in the same fashion as in \cite{pr}.
\subsection{Bifurcation of odd-periodic solutions.}\label{opp}\rm
Assume $\lambda_c$ denotes the first eigenvalues of $(I+\Delta)^2:
H_1 \hookrightarrow H$. Obviously $\lambda_c=\inf
\{P(\frac{\pi}{L}K)| K\in \mathbb{Z}^n \}$, where $P(x) =
(1-|x|^2)^2$, $x \in \mathbb{R}^n$.
{\theorem \label{abshop} The
following assertions are true for
the SH equation (\ref{SH-p}) with the odd periodic boundary condition :
\begin{enumerate}
\item For $\lambda \le \lambda_c$, $u=0$ is globally asymptotically stable.
\item for $\lambda > \lambda_c$, the SH equation (\ref{SH-p}) bifurcates from
$(u,\lambda)=(0,\lambda_c)$ to an
attractor $\mathcal{A}_\lambda$ which is homologic to $S^{n-1}$
(if $n=2$, $\mathcal{A}_\lambda$ is homeomorphic
$S^1$).
\item The attractor $\mathcal{A}_\lambda$ contains exactly $2^n$
steady state solutions of (\ref{SH-p}),
which are regular.
\item There is an $\varepsilon>0$ such that as
$\lambda_0<\lambda<\lambda_0 +\varepsilon$, the attractor $\mathcal
A_\lambda$ bifurcated from $(0,\lambda_0)$ attracts all bounded
sets in $H/ \Gamma$ in $H$, where $\Gamma$ is the stable manifold of
$u=0$ with codimension $n$.
\end{enumerate}}
{\proof Without loss of generality, we assume $L \le 2\pi$; so we
have
$\lambda_c = \lambda_1=(1-(\frac{2\pi}{L})^2)^2$. The eigenvalues and vectors
of $L_{\lambda}$ are given by the following:
\begin{equation}\label{evalue-op}
\begin{aligned}
&\beta_K(\lambda) = \lambda - (1- (\frac{2\pi}{L})^2|K|^2)^2,\\
&\phi_K(x) = \sqrt{\frac{2}{L^n}}sin(\frac{2\pi}{L}K.x),
\end{aligned}
\end{equation}
where $x=(x_1,x_2,...,x_n)$, and $ K=(k_1,k_2,....,k_n)$.

Now let $\beta_1(\lambda) = \beta_K(\lambda)$ and $ \phi_i = \phi_K$
when $K = (\delta_{i1},...., \delta_{in})$. Then we have the
following properties:
\begin{displaymath}
\begin{aligned}
&\beta_1(\lambda) \left\{%
\begin{array}{ccc}
<0 & if & \lambda<\lambda_1, \\
=0& if & \lambda=\lambda_1,\\
>0 & if & \lambda>\lambda_1,\\
\end{array}%
\right.\\ &\beta_K(\lambda_1) \le 0 \qquad \forall |K| \ge 2.
\end{aligned}
\end{displaymath}
So the first part of the theorem follows from the attractor
bifurcation theorem (\ref{abt}). In order to prove the second part
of the theorem we use the Lyapunov-Schmidt reduction method; this
gives us the following bifurcation equations:
\begin{equation}\label{4.8}
\beta_1(\lambda)y_i - \frac{3}{2L^2}(y_i ^3+ 2 \sum_{j\neq i}y_j ^2
y_i)=0.
\end{equation}
where $ 1\le i \le n$ and $y_i = y_K$ with $K = (\delta_{i1},....,
\delta_{in})$.

This equation has $2^n$ solutions as follows
\begin{equation}\label{4.9}
|y_1|=|y_2|=...=|y_n|= \cdots
= \sqrt{\frac{2L^2\beta_1(\lambda)}{3(2n-1)}}.
\end{equation}

It is easy to show that these solutions are regular.

For the case $n=2$, we have ${\mathcal A_{\lambda}} \simeq S^1$.
This follows from a similar reduction to the center manifold and
theorem \ref{s1b}. The circle in Figure4 depicts the center
manifold. The steady state points $y_1$ and $y_3$ are minimal
attractors and $y_2$ and $y_4$ are saddle points.

\begin{flushright}
\vspace{-.8cm} $\Box$
\end{flushright}
}
\begin{center}
\setlength{\unitlength}{1mm}
\begin{picture}(58,65)\label{lastFig}
\thicklines
\qbezier(25.000,10.000)(33.284,10.000)
(39.142,15.858)
\qbezier(39.142,15.858)(45.000,21.716)
(45.000,30.000)
\qbezier(45.000,30.000)(45.000,38.284)
(39.142,44.142)
\qbezier(39.142,44.142)(33.284,50.000)
(25.000,50.000)
\qbezier(25.000,50.000)(16.716,50.000)
(10.858,44.142)
\qbezier(10.858,44.142)( 5.000,38.284)
( 5.000,30.000)
\qbezier( 5.000,30.000)( 5.000,21.716)
(10.858,15.858)
\qbezier(10.858,15.858)(16.716,10.000)
(25.000,10.000)
\put(25.000,10.000){\vector(-1,0){0.5}}
\put(45.000,30.000){\vector(0,1){0.5}}
\put(25.000,50.000){\vector(1,0){0.5}}
\put( 5.000,30.000){\vector(0,-1){0.5}}

\put(25, 30){\vector(1,1){13}}
\put(25, 30){\vector(1,-1){13}}
\put(25, 30){\vector(-1,1){13}}
\put(25, 30){\vector(-1,-1){13}}
\put(25, 30){\line(1, 1){19}}
\put(25, 30){\line(-1, 1){19}}
\put(25, 30){\line(1, -1){19}}
\put(25, 30){\line(-1, -1){19}}
\put(41.142,13.858){\vector(-1,1){0.5}}
\put(41.142,46.142){\vector(-1,-1){0.5}}
\put(8.858,46.142){\vector(1,-1){0.5}}
\put(8.858,13.858){\vector(1,1){0.5}}
\put(39.142,15.858){\circle*{1}}
\put (42.142,15.858){$y_4$}
\put(39.142,44.142){\circle*{1}}
\put (38.142,48.142){$y_1$}
\put(10.858,44.142){\circle*{1}}
\put (4.858,43.142){$y_2$}
\put(10.858,15.858){\circle*{1}}
\put (10.858,10.858){$y_3$}
\end{picture}
\end{center}
\begin{center}
Figure4\\
\end{center}
\subsection{Bifurcation Of Periodic Solutions}\label{bp}
For the bifurcation of periodic solutions, the multiplicity of the
first eigenvalue is $2n$. Hence the long time behavior of the
bifurcated solutions will be essentially different from the previous
case.

\theorem \label{abshp} {The following assertions are true in the case of
the SH equation (\ref{SH-p}) with the periodic boundary condition :
\begin{enumerate}
\item For $\lambda \le \lambda_c$, $u=0$ is globally asymptotically stable.
\item for $\lambda > \lambda_c$, the equation bifurcates from $(u,\lambda)=(0,\lambda_c)$ to an
attractor $\mathcal{A}_\lambda$ which is homologic to
$S^{2n-1}$. Moreover, when $n=1$, $\mathcal{A}_\lambda$ is
homeomorphic to $S^1$.
\item The attractor $\mathcal{A}_\lambda$ contains an $n$-dimensional torus
$\mathbb{T}^{n}$, which consist of steady state solutions of (\ref{SH-p}).
\item There is an $\varepsilon>0$ such that as
$\lambda_c<\lambda<\lambda_c +\varepsilon$, the attractor $\mathcal
A_\lambda$ bifurcated from $(0,\lambda_c)$ attracts all bounded
sets in $H/ \Gamma$ in $H$, where $\Gamma$ is the stable manifold of
$u=0$ with codimension $2n$.

\end{enumerate}}
\proof{ \rm Without loss of generality, we assume $L \le 2\pi$; so
we have
$\lambda_c = \lambda_1=(1-(\frac{2\pi}{L})^2)^2$.
We get the eigenvectors and eigenvalues of $L_{\lambda} : H_1 \hookrightarrow H$ as follows:

\begin{equation}\label{evalue-p}
\begin{aligned}
&\beta_K(\lambda) = \lambda - (1- (\frac{2\pi}{L})^2|K|^2)^2,\\
&\phi_K(x) = \sqrt{\frac{2}{L^n}}sin(\frac{2\pi}{L}K.x),\\
&\psi_K (x) = \sqrt{\frac{2}{L^n}}cos(\frac{2\pi}{L}K.x).
\end{aligned}
\end{equation}
where $x=(x_1,x_2,...,x_n)$, and $ K=(k_1,k_2,....,k_n)$.\\

Now assume $\beta_1(\lambda) = \beta_K(\lambda)$, $ \phi_i =
\phi_K$, and $\psi_i = \psi_K$, when $K = (\delta_{i1},....,
\delta_{in})$. Then we have the following properties:
\begin{displaymath}
\begin{aligned}
&\beta_1(\lambda) \left\{%
\begin{array}{ccc}
<0 & if & \lambda<\lambda_1, \\
=0& if & \lambda=\lambda_1,\\
>0 & if & \lambda>\lambda_1,\\
\end{array}%
\right.\\ &\beta_K(\lambda_1) \le 0 \qquad \forall |K| \ge 2.
\end{aligned}
\end{displaymath}
So by Attractor Bifurcation Theorem (\ref{abt}), the first assertion
is true.

Assume $$ u = \sum_{i=1}^{i=n} (y_i\phi_i + z_i \psi_i) + \sum_{|K|>1}^{\infty}y_k \phi_K(x) + z_K \psi(x).$$
After reducing the equation SH to its center manifold, we get:

\begin{equation}\label{}
\begin{aligned}
&\frac{dy_i}{dt}= \beta_1(\lambda)y_i - \frac{3}{2L^2} y_i\sum_{j=1}^{j=m}(y_j ^2 + 2z_j ^2),\\
&\frac{dz_i}{dt}= \beta_1(\lambda)z_i - \frac{3}{2L^2}
z_i\sum_{j=1}^{j=m}(z_j ^2 + 2y_j ^2).\\
\end{aligned}
\end{equation}
for $ 1 \le i \le n$. This together with theorem (\ref{s1b}) show
that $\mathcal{A}_\lambda =S^1$ in the topological sense, when
$n=1$.

Since the subspace of the odd functions is an invariant set of $L_{\lambda} + G$ in $H$, the Lyapunov-Schmidt
reduction equations of the SH equation in this subspace are the same as
\ref{4.8}. Therefore the problem has solutions given in \ref{4.9}. Since the equation is invariant
under the spatial translation, steady state solutions associated with \ref{4.9} generate an
$n$-dimensional torus $\mathbb{T}^{n}$ as follows
$$\mathbb{T}^n = \{ \sqrt{\frac{2}{L^n}}\sum_{j=1}^{n} y_j sin(\frac{2\pi}{L}x_j + \theta_j) + o(|y|)~|
~~\forall ~(\theta_{1},..., \theta_{n}) \in \mathbb{R}^{n} \},$$
where $y = (y_1,...,y_n)$. This completes the proof.
}\begin{flushright}
\vspace{-.8cm} $\Box$
\end{flushright}

\section*{Acknowledgements} I would like to thank the
referees very much for their valuable comments and suggestions.

\medskip

Received May 2006; revised October 2006.

\medskip

\end{document}